\documentclass[12pt]{article}
\usepackage{amsmath,scalerel}
\usepackage{amsthm}
\usepackage{enumerate}
\usepackage{enumitem}
\usepackage{url}
\usepackage{tikz-cd}
\usepackage{mathtools}
\usepackage{comment}

\usepackage{amssymb}
\usepackage{graphics}
\graphicspath{{./}{Figures/}}

\usepackage{float}
\usepackage{color}
\usepackage{placeins}

\usepackage{hyperref}
\hypersetup{
    colorlinks=true, 
    linktoc=all,     
    linkcolor=blue,  
    linktocpage
}

\def\dsp{\def\baselinestretch{1.5}\large\normalsize}
\dsp
\allowdisplaybreaks
\begin{document}
\title{Accurate Computation of Activated Volume in Electromagnetic Heating}
%
\author{
Hongyun Wang\footnote{Corresponding author, hongwang@ucsc.edu} 
\\
Department of Applied Mathematics \\
University of California, Santa Cruz, CA 95064, USA\\
\\
Parthiv Seetharaman \\
Department of Mathematics \\
California Polytechnic State University, San Luis Obispo, CA 93407, USA\\
\\
Shannon E. Foley \\
U.S. Department of Defense \\
Joint Intermediate Force Capabilities Office \\
Quantico, VA 22134, USA\\
\\
Hong Zhou \\ Department of Applied Mathematics\\
Naval Postgraduate School, Monterey, CA 93943, USA
}
\maketitle

\newpage
\begin{abstract}
In electromagnetic heating and other applications, we need to compute the volume enclosed by an isosurface 
 of the 3D temperature distribution that is numerically represented on a rectangular grid. This situation 
 arises naturally when the temperature distribution is obtained by solving a partial differential equation numerically using a finite difference method. Given the temperature distribution on the 3D grid, the isosurface of a prescribed value is represented approximately by a triangulation, 
 a collection of triangles with vertices on the grid lines. The vertices are determined by a linear interpolation to approximate the locations where the temperature is at the prescribed level. 
 The region enclosed by the isosurface is approximated by that enclosed by the triangulation, which 
 is a set of tetrahedrons. The enclosed volume is approximated by summing those of tetrahedrons. 
 This volume approximation is analogous to approximating a curve using line segments
 and is limited in accuracy. 
 In this study, we combine extrapolation with the triangulation approximation 
 to develop a more accurate method for computing the volume enclosed 
 by an isosurface of the 3D temperature distribution on a rectangular grid. 
%
\end{abstract}

\noindent\textbf{Keywords:} isosurface, volume computation, triangulation, extrapolation, rectangular grid

\section{Introduction} 
In electromagnetic heating of human body, skin surface is exposed to a millimeter waves (MMW)
in the frequency range of 30–300 GHz. The millimeter-wave band corresponds to wavelengths 
between 1 mm at 300 GHz and 10 mm at 30 GHz.
MMWs have a wide range of applications \cite{Romanenko_2017, Zhadobov_2011,
Quement_2014, Nelson_2000, Foster_2010, Cazares_2019, Topfer_2015},
including 5G networks \cite{Sri_2020}, detection of explosives \cite{Sheen_2007}, 
and autonomous railway systems \cite{Liu_2020}. Given their increasing prevalence, it is important to study the biological effects of MMW exposure on humans. 
The primary effect of MMW exposure on humans is the electromagnetic heating. 
Skin is a lossy medium. MMW wave is rapidly attenuated during propagation in skin 
and the electromagnetic power is absorbed within a thin layer of skin, 
which becomes a heat source in the skin thermal evolution. 
When the local temperature reaches the activation threshold, thermal nociceptors at that skin
location are activated, inducing a heating sensation. 
Over an extended duration of MMW exposure, the skin temperature keeps increasing, 
which activates nociceptors over a larger region and intensifies the induced heating sensation. 
For the purpose of assessing the induced heating sensation and/or assessing the thermal injury risk, 
it is necessary to compute the skin volume where the local temperature is at or 
above a given value. 
This skin volume corresponds to the region enclosed by the isosurface of 
the 3D skin temperature at the given level. In this study, we cast this task into a broad mathematical 
setting and consider the general problem of computing the 3D volume enclosed by 
an isosurface of function $f(x, y, z)$. 
We focus on the practical situation where function $f(x, y, z)$ is described only discretely on a 
3D rectangular grid. This situation arises naturally in many thermal applications where 
the temperature distribution is obtained by solving the heat equation numerically on a 3D grid. 

The remainder of the paper is organized as follows. Section 2 presents the mathematical formulation 
of the method for computing the enclosed volume and for carrying out extrapolation. 
Section 3 examines the accuracy of the method, respectively, without and with extrapolation, 
on several test problems in which the exact enclosed volume is known analytically. 
Section 4 tests the performance of the methods on a simplified case of electromagnetic heating
where an idealized skin tissue of uniform material properties is exposed to 
an axial-symmetric Gaussian beam at perpendicular incident and the lateral heat conduction 
is neglected. In this idealized electromagnetic heating problem, 
the activated volume can be practically calculated up to machine precision by utilizing the particular 
axial symmetry in this model problem. 
Section 5 summarizes the main results of the study.

\section{Mathematical Formulation}
Consider region $D$ in $\mathbb{R}^3$. We view it as enclosed by its boundary surface 
$S = \partial D$. 
Let $\mathbf{x}_0 = (x_0, y_0, z_0)$ be a point in $\mathbb{R}^3$. We use 
the divergence theorem and the fact that $\nabla \cdot (\mathbf{x} - \mathbf{x}_0) = 3$
to connect a surface integral to the volume of region $D$.
\begin{equation}
\underbrace{\oint_{\partial D} (\mathbf{x} - \mathbf{x}_0) \cdot  \mathbf{dS} 
=  \int_{D} \nabla \cdot (\mathbf{x} - \mathbf{x}_0) dV}_{\text{divergence theorem}} 
 = 3 \times \text{volume}(D) 
 \label{diverg_thm}
\end{equation}
where $\mathbf{dS} = \hat{\mathbf{n}}\, dS $ is the vector area element, containing 
both the magnitude of the area element $dS$
and its outward unit normal direction $\hat{\mathbf{n}}$. 
From \eqref{diverg_thm}, we express the volume of region $D$ in terms of 
a surface integral over its boundary surface. 
\begin{equation}
\text{volume}(D) = \frac{1}{3} \oint_{S}
(\mathbf{x} - \mathbf{x}_0) \cdot  \mathbf{dS}, \qquad \mathbf{x}_0 \in \mathbb{R}^3
\label{V_act}
\end{equation}
where $S = \partial D$ is the enclosing boundary surface. 
\eqref{V_act} is valid no matter whether $\mathbf{x}_0$ is inside or outside region $D$ 
or on its boundary surface. 

In electromagnetic heating applications, region $D$ is defined as having temperature at or 
above a given level. As a result, its boundary surface $S = \partial D$ is the isosurface 
of the 3D skin temperature at the given level. 
Let $f(\mathbf{x})$ be the skin temperature at position $\mathbf{x}$. We consider the situation
where function $f(\mathbf{x})$ is only represented by its values on a 3D rectangular grid. 
In standard computational packages (such as MatLab), 
given a discrete representation of function $f(\mathbf{x})$ and a level value, 
the isosurface $S$ is approximated by a triangulation representation, 
a collection of triangular patches. 
\begin{equation}
S = \sum_{j=1}^N S_j \approx \sum_{j=1}^N \underbrace{\quad 
\Delta(\mathbf{x}_1^{(j)}, \mathbf{x}_2^{(j)}, \mathbf{x}_3^{(j)}) \quad }_{\text{planar patch approximation of $S_j$}}
\label{S_approx}
\end{equation} 
In \eqref{S_approx}, $S_j$ represents a small piece of the true curly surface $S$ while 
$\Delta(\mathbf{x}_1^{(j)}, \mathbf{x}_2^{(j)}, \mathbf{x}_3^{(j)})$ 
is a triangular patch approximating $S_j$. $N$ is the number of patches in the triangulation.
The surface integral over $S$ is approximated by the sum of integrals over individual 
patches.
\begin{align}
& \oint_{S} (\mathbf{x} - \mathbf{x}_0) \cdot  \mathbf{dS} \approx  \sum_{j=1}^N 
\int_{\Delta(\mathbf{x}_1^{(j)}, \mathbf{x}_2^{(j)}, \mathbf{x}_3^{(j)})} (\mathbf{x} - \mathbf{x}_0) \cdot  \mathbf{dS} 
\label{sur_int}
\end{align}
Each patch is a planar triangle. The integral over a patch has an analytical expression
\begin{equation}
\int_{\Delta(\mathbf{x}_1^{(j)}, \mathbf{x}_2^{(j)}, \mathbf{x}_3^{(j)})} 
(\mathbf{x} - \mathbf{x}_0) \cdot  \mathbf{dS}
= \frac{1}{2}\Big((\mathbf{x}_1^{(j)}-\mathbf{x}_0) \times (\mathbf{x}_2^{(j)}-\mathbf{x}_0) \Big)
\cdot (\mathbf{x}_3^{(j)}-\mathbf{x}_0) 
\label{tetrah}
\end{equation}
\begin{figure}[!h]
\vskip 0.0cm
\begin{center}
\includegraphics[width=3.0in]{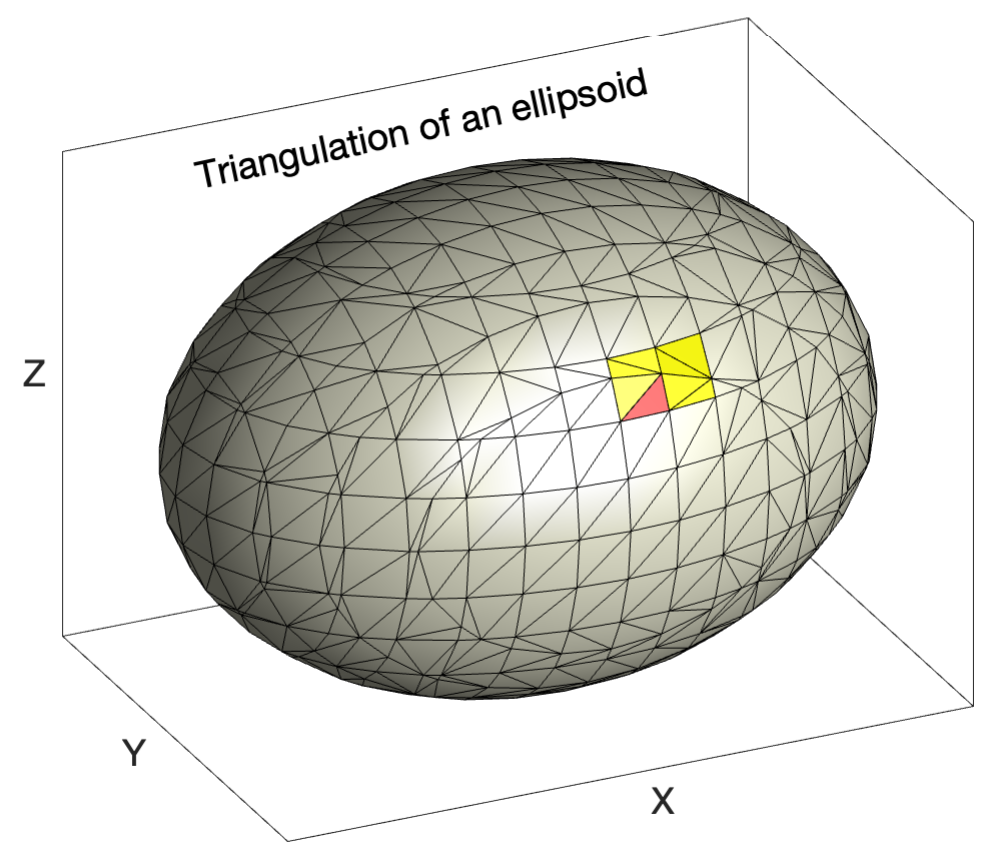} \\
\includegraphics[width=3.0in]{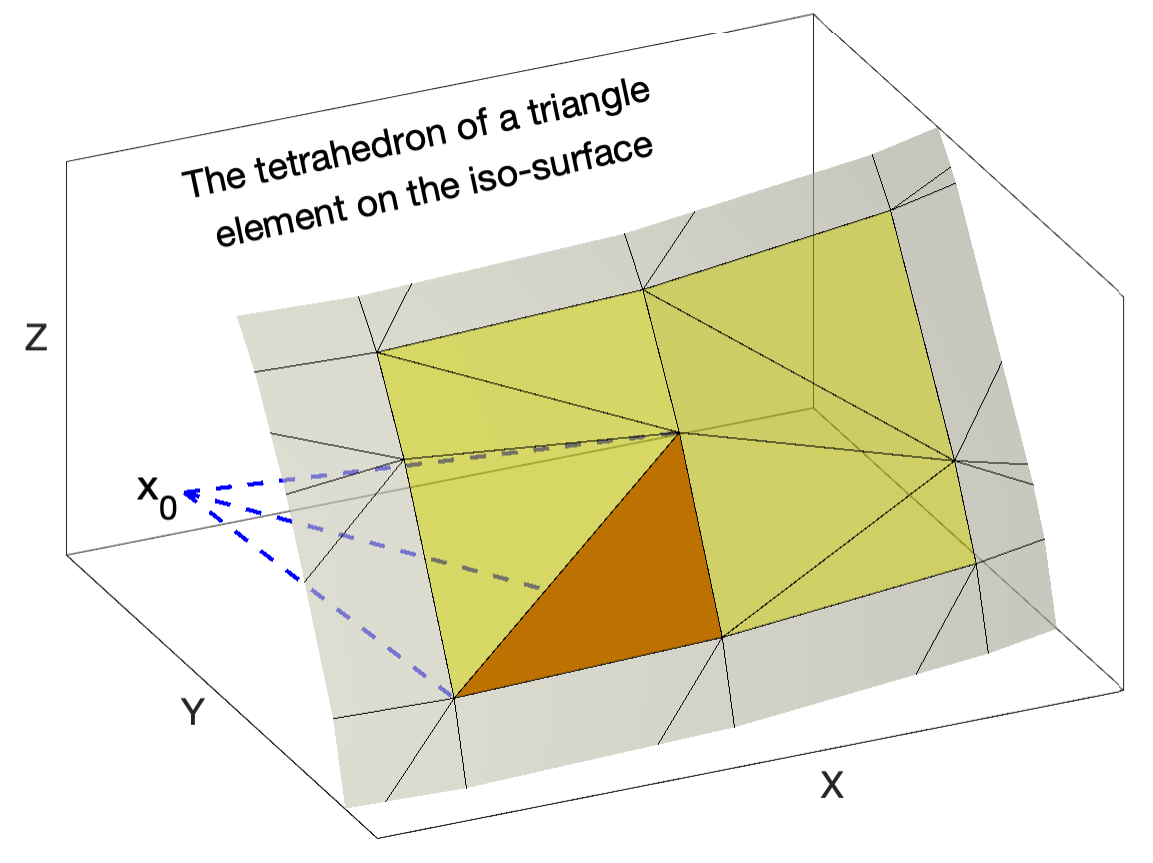}
\end{center}
\vskip -0.75cm
\caption{Top: region enclosed by the triangulation of an isosurface. 
Bottom: detailed view of a triangle element in the isosurface triangulation.} 
\label{fig_method}
\end{figure}
Therefore, the volume of region $D$ (the region enclosed by surface $S$) is approximated by 
\begin{equation}
\boxed{\quad 
\text{volume}(D) \approx \underbrace{\frac{1}{6}\sum_{j=1}^N 
\Big((\mathbf{x}_1^{(j)}-\mathbf{x}_0) \times (\mathbf{x}_2^{(j)}-\mathbf{x}_0) \Big)
\cdot (\mathbf{x}_3^{(j)}-\mathbf{x}_0)}_{\text{volume enclosed by the triangulation}} 
\equiv V(\Delta x) \quad}
\label{V_method}
\end{equation}
where $V(\Delta x)$ denotes the volume enclosed by the triangulation of the isosurface 
based on the discrete representation of function $f(\mathbf{x})$ 
on the grid of mesh spacing $\Delta x$. 
$V(\Delta x)$ is used as an approximation to the volume enclosed by the true isosurface, 
which is not known exactly and has some uncertainty if function $f(\mathbf{x})$ is 
represented only on a discrete grid. 
In particular, the notation $V(\Delta x)$ indicates explicitly the effect of 
mesh spacing $\Delta x$ on the iso-surface triangulation and on the volume approximation. 

In the triangulation, the true isosurface in the 3D is replaced with a set of planar triangle elements.
This is analogous to the situation where a curve in the 3D is replace with a set of line segments, 
which yields a second order approximation. 
Thus, we expect $V(\Delta x) $ to be a second order approximation of the true volume 
$ V_\text{true} \equiv \text{volume}(D)$. 
\begin{equation}
V(\Delta x) = V_\text{true} + C_2 (\Delta x)^2 + o\big( (\Delta x)^2\big) 
\label{V_dx}
\end{equation}
Given the discrete representation of function $f(\mathbf{x})$ on a 3D grid of mesh spacing 
$\Delta x$, we can alway down-sample to obtain a coarse representation of function 
$f(\mathbf{x})$ on a grid of mesh spacing $(2\Delta x)$. From the coarse representation, 
we construct a coarse triangulation of the isosurface and the corresponding volume approximation 
$ V(2\Delta x)$, which satisfies 
\begin{equation}
V(2\Delta x) = V_\text{true} + C_2 (2\Delta x)^2 + o\big( (\Delta x)^2\big) 
\label{V_2dx}
\end{equation}
In \eqref{V_dx}, the dominant part of error in $V(\Delta x)$ is $C_2(\Delta x)^2$. 
\eqref{V_2dx} based on the coarse representation contains a similar term $C_2(2 \Delta x)^2$. 
We combine \eqref{V_dx} and \eqref{V_2dx} to eliminate the dominant error 
$C_2(\Delta x)^2$ in $V(\Delta x)$. 
The result is a more accurate approximation to $ V_\text{true}$. 
\begin{equation}
\boxed{\quad V_\text{extrap}(\Delta x) \equiv \frac{4 V(\Delta x) - V(2\Delta x)}{3} 
= V_\text{true} +  o\big( (\Delta x)^2\big) \quad }
\label{V_extrap}
\end{equation}
Mathematically, this is an extrapolation: we are using $V(2\Delta x)$ and $V(\Delta x)$ 
to predict $V(0_+)$, the numerical approximation at an infinitesimal mesh spacing. 

\section{Several test problems of enclosed volumes with analytical solutions}
We study the numerical accuracy of the volume computation method \eqref{V_method} and 
the extrapolation improvement \eqref{V_extrap}. 
To examine the true numerical error, we select several test problems in which 
the volume enclosed by the given iso-surface has a closed-form analytical expression. 
In the discussion below, let $f(\mathbf{x})$ be the underlying function, mimicking the role 
of the 3D skin temperature distribution in the electromagnetic heating problem. 
Let $D$ be the region enclosed by the isosurface of function $f(\mathbf{x})$ at 
a given value $f_c$. As introduced in the last section, $\text{volume}(D)$ 
denotes the true volume of the enclosed region $D$, and 
$V(\Delta x)$ and $V_\text{extrap}(\Delta x)$ denote the numerical volumes 
of respectively method \eqref{V_method} and method \eqref{V_extrap}
based on a discrete representation of $f(\mathbf{x})$ on a grid of mesh spacing $\Delta x$. 
Method \eqref{V_extrap} is built on method \eqref{V_method} with extrapolation 
incorporated. 

The true numerical errors respectively in $V(\Delta x)$ and in $V_\text{extrap}(\Delta x)$ 
are defined as 
\begin{equation}
\begin{dcases}
 \big(\text{error in }V(\Delta x) \big) = \big| V(\Delta x) - \text{volume}(D) \big| \\[1ex] 
 \big(\text{error in }V_\text{extrap}(\Delta x) \big) = \big| V_\text{extrap}(\Delta x) - 
\text{volume}(D) \big| 
\end{dcases}
\label{num_err}
\end{equation}
\subsection{The unit sphere} \label{P0}
\begin{equation}
\begin{dcases}
\text{function: } f(\mathbf{x}) = 1-(x^2 + y^2 + z^2), \quad  \mathbf{x} = (x, y, z) \\
\text{enclosed region: } D = \{(x, y, z) \big| f(\mathbf{x}) \ge 0 \} \\[1ex]
\text{volume}(D) = \frac{4 \pi}{3}
\end{dcases}
\label{prob_0}
\end{equation}
\begin{figure}[!h]
\vskip 0.0cm
\begin{center}
\includegraphics[height=2.6in]{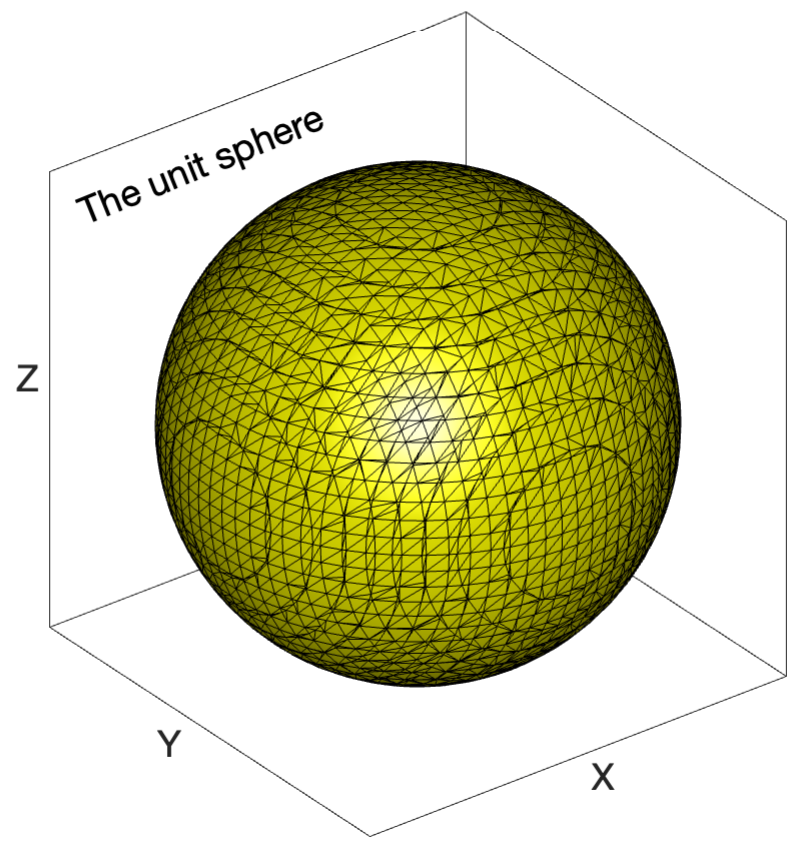} \;\;
\includegraphics[width=3.2in]{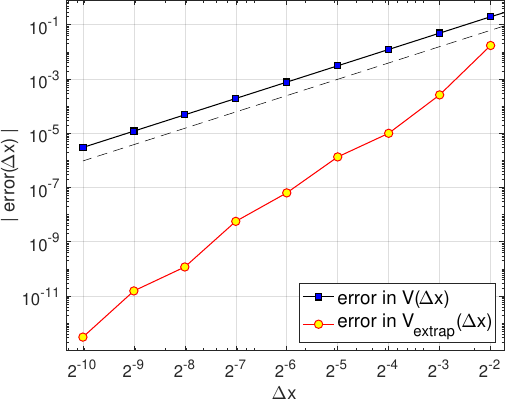}
\end{center}
\vskip -0.75cm
\caption{Left: Triangulation of the unit sphere as an isosurface of function $f(\mathbf{x})$
given in \eqref{prob_0}. Right: errors in numerical volumes $V(\Delta x)$ and 
$V_\text{extrap}(\Delta x)$ vs mesh spacing $\Delta x$. Dashed line 
is $(\Delta x)^2$ vs $\Delta x$ serving as a visual guide.} 
\label{fig_00}
\end{figure}
Numerical results for the unit sphere are shown in Figure \ref{fig_00}. 
As predicted in \eqref{V_dx}, the error in $V(\Delta x)$ is approximately 
proportional to $(\Delta x)^2$. The error in $V_\text{extrap}(\Delta x)$ is significantly 
below that in $V(\Delta x)$, demonstrating the benefit of extrapolation. Note that
the extrapolation component does not require addition data or information on function 
$f(\mathbf{x})$. $V(2\Delta x)$ in the extrapolation is obtained from down sampling the 
given discrete representation of $f(\mathbf{x})$. 

\subsection{An axis-aligned ellipsoid} \label{P1}
\begin{equation}
\begin{dcases}
\text{function: } f(\mathbf{x}) = 1-(\frac{x^2}{a^2} + \frac{y^2}{b^2} +\frac{z^2}{c^2}), \quad  \mathbf{x} = (x, y, z) \\[0ex]
\qquad \text{where } \;\; a = 1, \;\; b = \frac{1}{4}, \;\; c = \frac{1}{2} \\[1ex]
\text{enclosed region: } D = \{(x, y, z) \big| f(\mathbf{x}) \ge 0 \} \\[1ex]
\text{volume}(D) = \frac{4 \pi}{3}a b c
\end{dcases}
\label{prob_1}
\end{equation}
\begin{figure}[!h]
\vskip 0.0cm
\begin{center}
\includegraphics[height=2.6in]{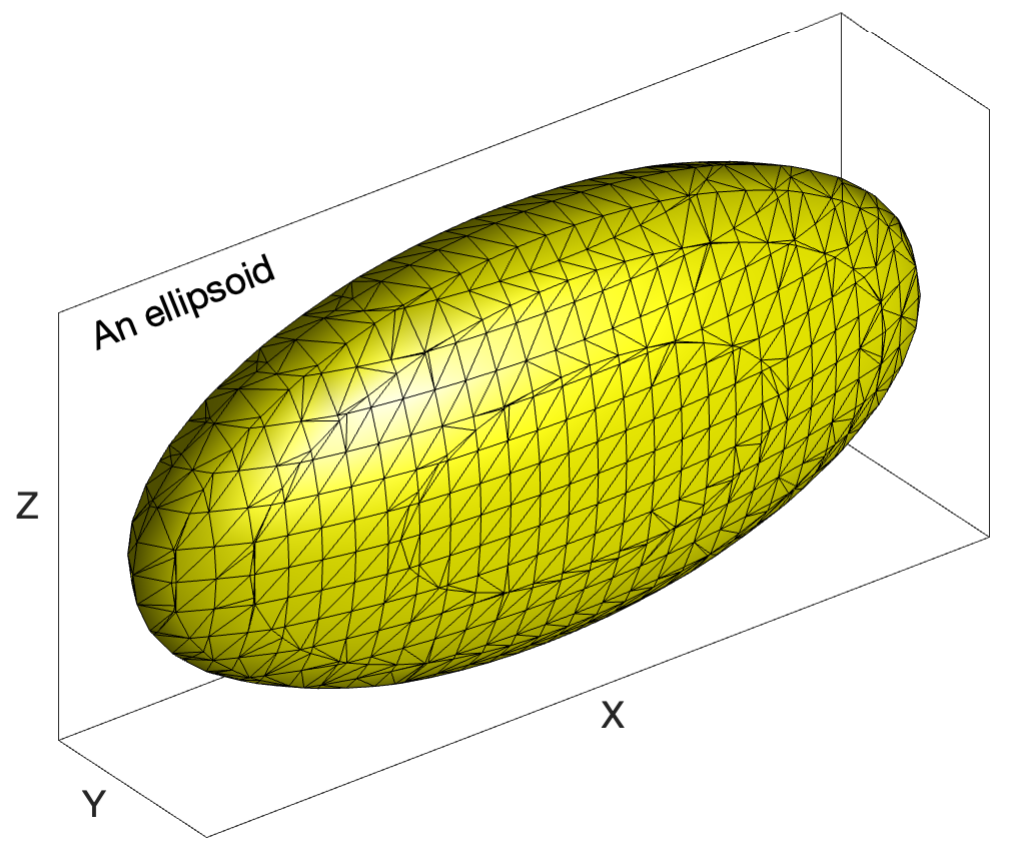} \;\;
\includegraphics[width=3.2in]{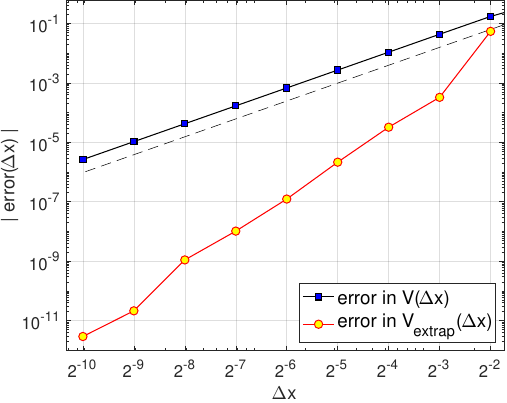}
\end{center}
\vskip -0.75cm
\caption{Left: Triangulation of the axis-aligned ellipsoid as an isosurface of 
$f(\mathbf{x})$ in \eqref{prob_1}. Right: errors in numerical volumes $V(\Delta x)$ and 
$V_\text{extrap}(\Delta x)$ vs mesh spacing $\Delta x$.} 
\label{fig_01}
\end{figure}
In Figure \ref{fig_01} for the axis-aligned ellipsoid , the volume computation methods show the same behavior as in Figure \ref{fig_00}. 
The error in $V(\Delta x)$ is approximately proportional to $(\Delta x)^2$; 
the error in $V_\text{extrap}(\Delta x)$ is significantly smaller than that in $V(\Delta x)$, 
confirming the second order accuracy of triangulation and the accuracy improvement 
of extrapolation. 

\subsection{An rotated ellipsoid} \label{P2}
\begin{equation}
\begin{dcases}
\text{function: } f(\mathbf{x}) = 1-(\frac{\tilde{x}^2}{a^2} +\frac{\tilde{y}^2}{b^2} 
+\frac{\tilde{z}^2}{c^2}), \quad  \mathbf{x} = (x, y, z) \\[0ex] 
\qquad \text{where }
\begin{pmatrix} \tilde{x} \\ \tilde{y} \\ \tilde{z} \end{pmatrix}
= A \begin{pmatrix} x \\ y \\ z \end{pmatrix}, \quad 
A = \begin{pmatrix} \frac{\sqrt{3}}{2\sqrt{2} } & \frac{1}{\sqrt{2}} & \frac{1}{2\sqrt{2}} \\[1ex]
\frac{-\sqrt{3}}{2\sqrt{2} } & \frac{1}{\sqrt{2}} & \frac{-1}{2\sqrt{2}} \\ 
\frac{-1}{2} & 0 & \frac{\sqrt{3}}{2} \end{pmatrix} \\[0ex]
\qquad \qquad a = 1, \;\; b = \frac{1}{4}, \;\; c = \frac{1}{2} \\[1ex]
\text{enclosed region: } D = \{(x, y, z) \big| f(\mathbf{x}) \ge 0 \} \\[1ex]
\text{volume}(D) = \frac{4 \pi}{3}a b c
\end{dcases}
\label{prob_2}
\end{equation}
In  \eqref{prob_2}, matrix $A$ is orthogonal, satisfying $A^T A = I$. Geometrically, multiplying by $A$
corresponds to rotating the 3D region with respect to the $z$-axis by $45^\circ$ and then rotating 
it with respect to the $y$-axis by $(-30^\circ)$. 
The enclosed region in \eqref{prob_2} is congruent to that of subsection \ref{P1} 
via an orthogonal transformation, which preserves the analytical volume. 
\begin{figure}[!h]
\vskip 0.0cm
\begin{center}
\includegraphics[height=2.6in]{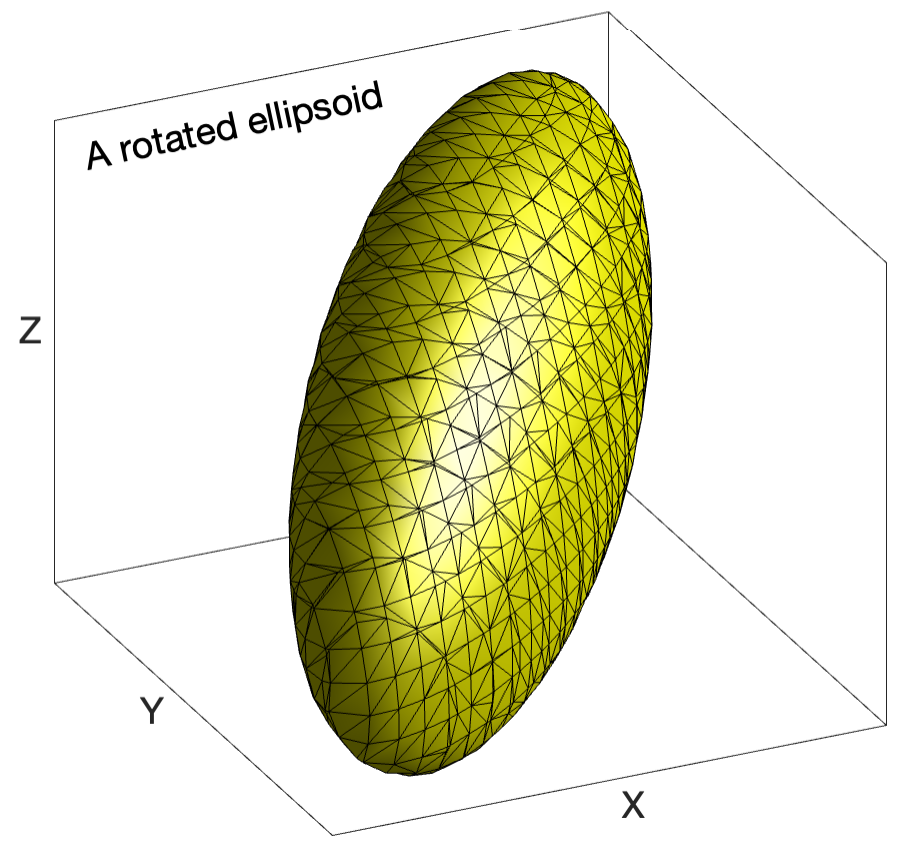} \;\;
\includegraphics[width=3.2in]{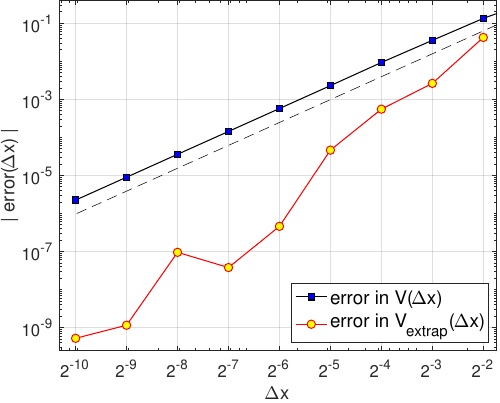}
\end{center}
\vskip -0.75cm
\caption{Left: Triangulation of the rotated ellipsoid as an isosurface of 
$f(\mathbf{x})$ in \eqref{prob_2}. Right: errors in numerical volumes $V(\Delta x)$ and 
$V_\text{extrap}(\Delta x)$ vs mesh spacing $\Delta x$.} 
\label{fig_02}
\end{figure}
The discrete representation of function $f(\mathbf{x})$ on a fixed 3D grid, however, 
varies with the orientation. Accordingly the corresponding 
numerical volume based on the discrete representation is expected to be affected. 
In this sense, the rotated ellipsoid provides a completely new test on the 
volume computation methods although the exact volume stays the same. 
For the rotated ellipsoid, in Figure \ref{fig_02} we observe the same behavior as in 
Figures \ref{fig_00}-\ref{fig_01}.
The error in $V(\Delta x)$ is approximately proportional to $(\Delta x)^2$; 
the error in $V_\text{extrap}(\Delta x)$ is much lower than that in $V(\Delta x)$. 

\subsection{An axis-aligned superellipsoid} \label{P3}
\begin{equation}
\begin{dcases}
\text{function: } f(\mathbf{x}) = 1-\Big(
\Big[\Big(\frac{x}{a}\Big)^{\frac{2}{\varepsilon_2}}
+\Big(\frac{y}{b}\Big)^{\frac{2}{\varepsilon_2}} \Big]^{\frac{\varepsilon_2}{\varepsilon_1}}
+ \Big(\frac{z}{c}\Big)^{\frac{2}{\varepsilon_1}}
\Big), \quad  \mathbf{x} = (x, y, z) \\[0ex] 
\qquad \text{where } \;\; \varepsilon_1 = \varepsilon_2 = \frac{3}{4}, \;\; 
a = 1, \;\; b = \frac{1}{4}, \;\; c = \frac{1}{2} \\[1ex]
\text{enclosed region: } D = \{(x, y, z) \big| f(\mathbf{x}) \ge 0 \} \\[1ex]
\text{volume}(D) = 2a b c\, \varepsilon_1\varepsilon_2 \, 
\beta(\frac{\varepsilon_1}{2}, \varepsilon_1 +1)
\beta(\frac{\varepsilon_2}{2}, \varepsilon_2 +2)
\end{dcases}
\label{prob_3}
\end{equation}
where $\beta(a, b) $ is the beta function defined as 
\[ \beta(a, b) \equiv \int_0^1 t^{a-1} (1-t)^{b-1} dt \] 
The beta function is available in standard software packages. 
\begin{figure}[!h]
\vskip 0.0cm
\begin{center}
\includegraphics[height=2.6in]{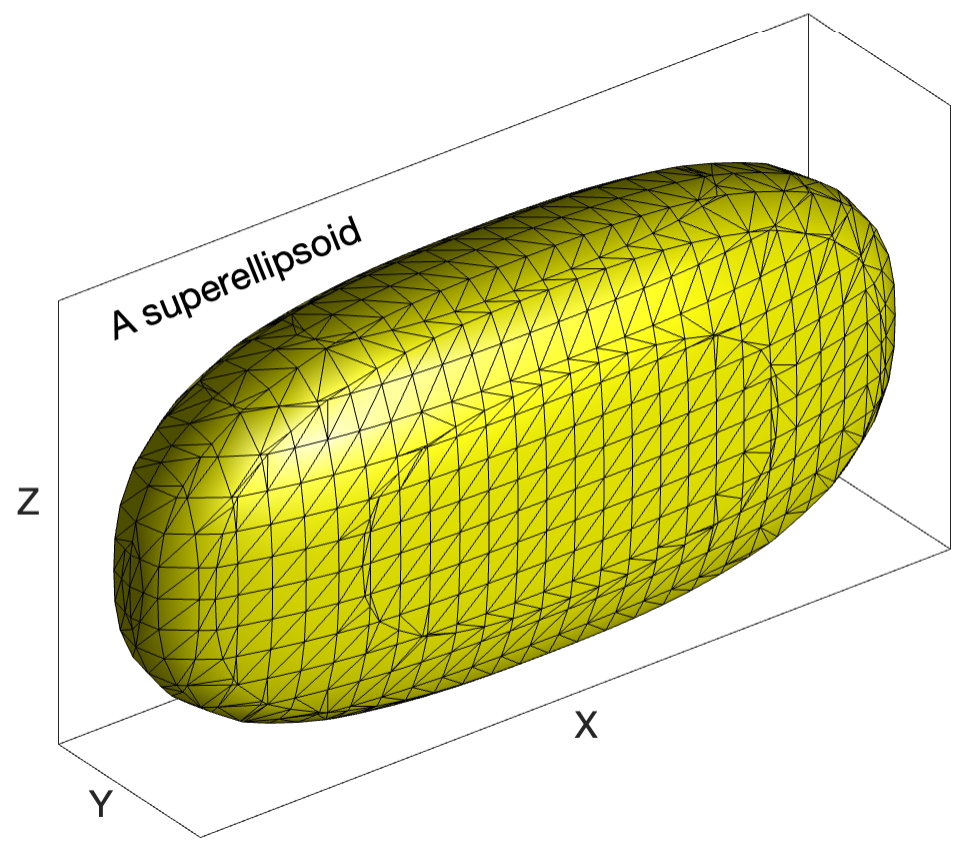} \;\;
\includegraphics[width=3.2in]{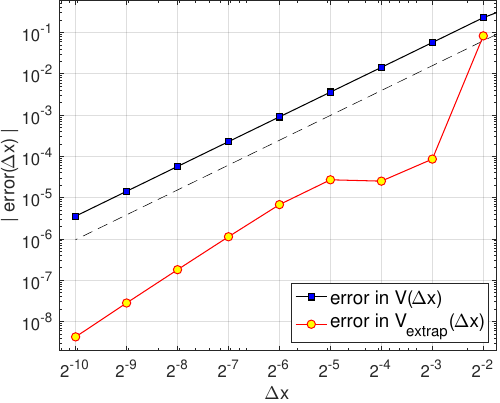}
\end{center}
\vskip -0.75cm
\caption{Left: Triangulation of the axis-aligned superellipsoid as an isosurface of 
$f(\mathbf{x})$ in \eqref{prob_3}. Right: errors in numerical volumes $V(\Delta x)$ and 
$V_\text{extrap}(\Delta x)$ vs mesh spacing $\Delta x$.} 
\label{fig_03}
\end{figure}
The regular ellipsoid \eqref{prob_1} is a special case of 
the superellipsoid \eqref{prob_3} at $\varepsilon_1 = \varepsilon_2 = 1$. 
In the case of $\varepsilon_1 = \varepsilon_2 = \frac{2}{n}$, 
the enclosed region is described by 
\[ \Big(\frac{x}{a}\Big)^n +\Big(\frac{y}{b}\Big)^n + 
\Big(\frac{z}{c}\Big)^n \le 1 \]
As $n \rightarrow +\infty$, the enclosed region converges to the 3D rectangular prism: 
\[ [-a, a] \times  [-b, b] \times [-c, c] \] 
For  $\varepsilon_1 = \varepsilon_2 < 1$, the enclosed superellipsoid is between 
the regular ellipsoid and the 3D rectangular prism. 
Indeed, the principal sides of the superellipsoid shown in the left panel of Figure \ref{fig_03} 
are flatter than those of the regular ellipsoid in Figure \ref{fig_01}.
For the superellipsoid, even with this shape change in the enclosed region, 
the error trends in volume computation remain the same as in Figures \ref{fig_00}-\ref{fig_02}. 
The error in $V(\Delta x)$ is proportional to $(\Delta x)^2$; the error in 
$V_\text{extrap}(\Delta x)$ is significantly smaller than that in $V(\Delta x)$. 

\subsection{A rotated superellipsoid} \label{P4}
\begin{equation}
\begin{dcases}
\text{function: } f(\mathbf{x}) = 1-\Big(
\Big[\Big(\frac{\tilde{x}}{a}\Big)^{\frac{2}{\varepsilon_2}}
+\Big(\frac{\tilde{y}}{b}\Big)^{\frac{2}{\varepsilon_2}} \Big]^{\frac{\varepsilon_2}{\varepsilon_1}}+ \Big(\frac{\tilde{z}}{c}\Big)^{\frac{2}{\varepsilon_1}}
\Big), \quad  \mathbf{x} = (x, y, z) \\[0ex] 
\qquad \text{where }
\begin{pmatrix} \tilde{x} \\ \tilde{y} \\ \tilde{z} \end{pmatrix}
= A \begin{pmatrix} x \\ y \\ z \end{pmatrix}, \quad 
A = \begin{pmatrix} \frac{\sqrt{3}}{2\sqrt{2} } & \frac{1}{\sqrt{2}} & \frac{1}{2\sqrt{2}} \\[1ex]
\frac{-\sqrt{3}}{2\sqrt{2} } & \frac{1}{\sqrt{2}} & \frac{-1}{2\sqrt{2}} \\ 
\frac{-1}{2} & 0 & \frac{\sqrt{3}}{2} \end{pmatrix} \\[0ex]
\qquad \qquad \varepsilon_1 = \varepsilon_2 = \frac{3}{4}, \;\; 
a = 1, \;\; b = \frac{1}{4}, \;\; c = \frac{1}{2} \\[1ex]
\text{enclosed region: } D = \{(x, y, z) \big| f(\mathbf{x}) \ge 0 \} \\[1ex]
\text{volume}(D) = 2a b c\, \varepsilon_1\varepsilon_2 \, 
\beta(\frac{\varepsilon_1}{2}, \varepsilon_1 +1)
\beta(\frac{\varepsilon_2}{2}, \varepsilon_2 +2)
\end{dcases}
\label{prob_4}
\end{equation}
The enclosed region in \eqref{prob_4} is congruent to that in \eqref{prob_3} of subsection \ref{P3} 
via an orthogonal transformation. In \eqref{prob_4}, 
the orthogonal transformation represented by matrix $A$ is the same as 
that in \eqref{prob_2}. 
Similar to what we observed in Figures \ref{fig_00}-\ref{fig_03}, 
the error in $V(\Delta x)$ is proportional to $(\Delta x)^2$; the error in 
$V_\text{extrap}(\Delta x)$ is significantly smaller than that in $V(\Delta x)$, 
demonstrating the steady improvement of extrapolation over several test problems. 
\begin{figure}[!h]
\vskip 0.0cm
\begin{center}
\includegraphics[height=2.6in]{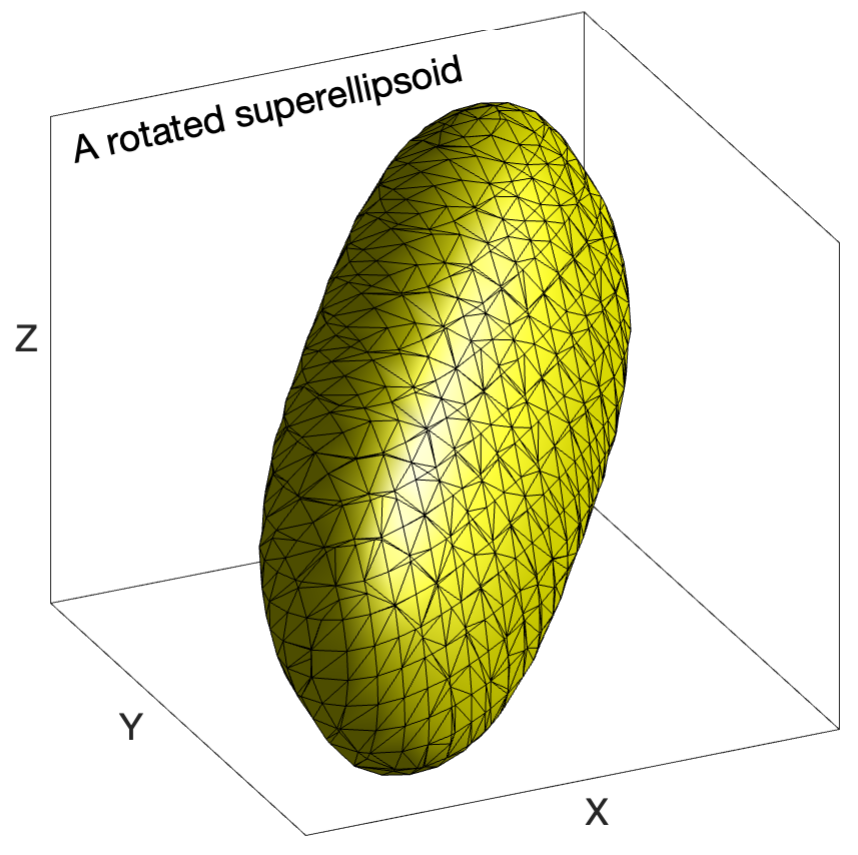} \;\;
\includegraphics[width=3.2in]{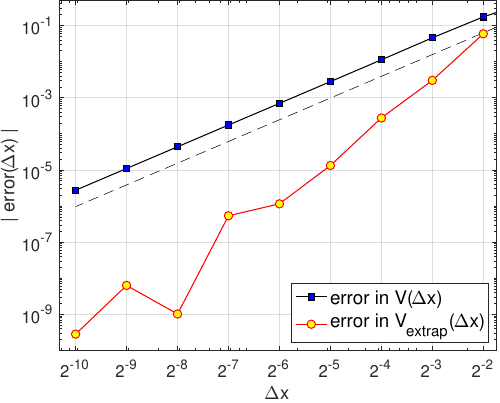}
\end{center}
\vskip -0.75cm 
\caption{Left: Triangulation of the rotated superellipsoid as an isosurface of 
$f(\mathbf{x})$ in \eqref{prob_4}. Right: errors in numerical volumes $V(\Delta x)$ and 
$V_\text{extrap}(\Delta x)$ vs mesh spacing $\Delta x$.} 
\label{fig_04}
\end{figure}
\FloatBarrier

\subsection{An axis-aligned elliptic ring torus}
\begin{figure}[!h]
\vskip 0.0cm
\begin{center}
\includegraphics[height=2.5in]{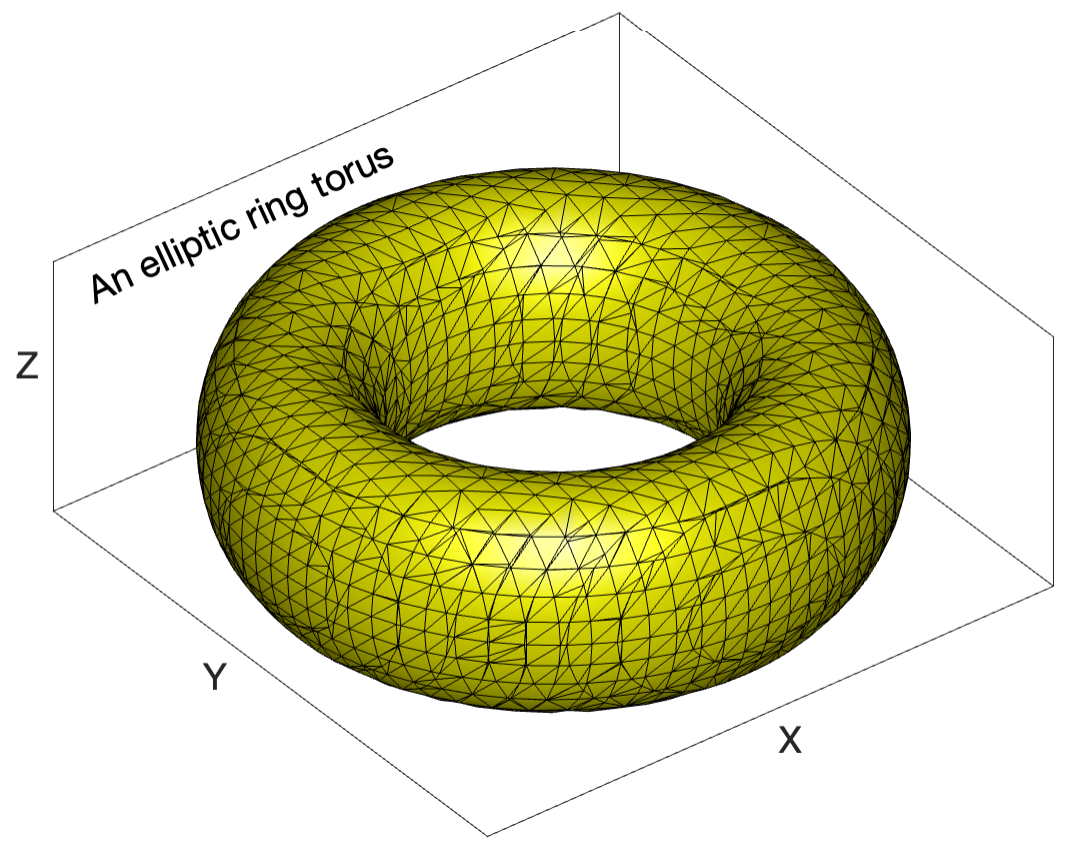} \;\;
\includegraphics[width=3.0in]{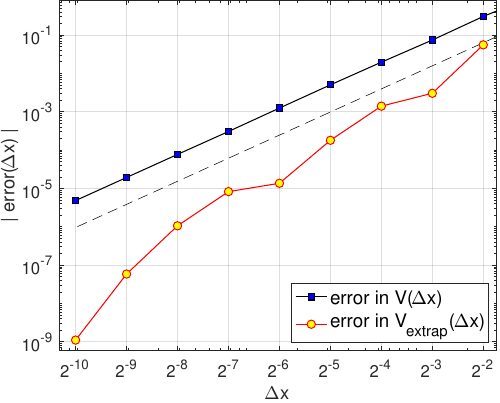}
\end{center}
\vskip -0.75cm
\caption{Left: Triangulation of the axis-aligned elliptic ring torus as an isosurface of 
$f(\mathbf{x})$ in \eqref{prob_5}. Right: errors in numerical volumes $V(\Delta x)$ and 
$V_\text{extrap}(\Delta x)$ vs mesh spacing $\Delta x$.} 
\label{fig_05}
\end{figure}

\begin{equation}
\begin{dcases}
\text{function: } f(\mathbf{x}) = r^2-\Big(
\Big[ \sqrt{x^2+y^2} - R\Big]^2+\Big(\frac{z}{c}\Big)^2 \Big), \quad  \mathbf{x} = (x, y, z) 
\\[0ex] 
\qquad \text{where } \;\; R = 0.8, \;\; r = 0.28, \;\; 
c = \frac{1}{0.7} \\[1ex]
\text{enclosed region: } D = \{(x, y, z) \big| f(\mathbf{x}) \ge 0 \} \\[1ex]
\text{volume}(D) = 2 \pi^2 c\, r^2 R  
\end{dcases}
\label{prob_5}
\end{equation}
A triangulation of the axis-aligned elliptic ring torus is shown in the left panel of Figure \ref{fig_05}. 
Geometrically, the centerline of the torus is the circle of radius $R$ centered at the origin 
in the $(x, y)$-plane. 
Perpendicular to the centerline, the cross section of the torus is an ellipse with 
the semi minor axis $r$ in the horizontal direction and 
the semi major axis $(c\, r)$ in the $z$-direction. 
For the elliptic ring torus in \eqref{prob_5}, $c > 1$. 
The exact volume of the enclosed region is given by the product of 
the center line arclength ($2 \pi R$) and the cross section area ($\pi c\, r^2$). 
We use the ring torus as a test problem to examine the error trends in numerical volumes. 
Note that numerical volumes are computed from a discrete representation of 
function $f(\mathbf{x})$ on a 3D grid of mesh spacing $\Delta x$, 
not using the full analytical expression of $f(\mathbf{x})$. 
The exact volume is not used in computing numerical volumes; 
it is used only in evaluating the errors of numerical volumes. 
The right panel of Figure \ref{fig_05} shows that the errors in 
$V(\Delta x)$ and $V_\text{extrap}(\Delta x)$ both decrease with $\Delta x$
with the latter significantly below the former. 
\FloatBarrier

\subsection{A rotated elliptic ring torus}
\begin{figure}[!h]
\vskip 0.0cm
\begin{center}
\includegraphics[height=2.5in]{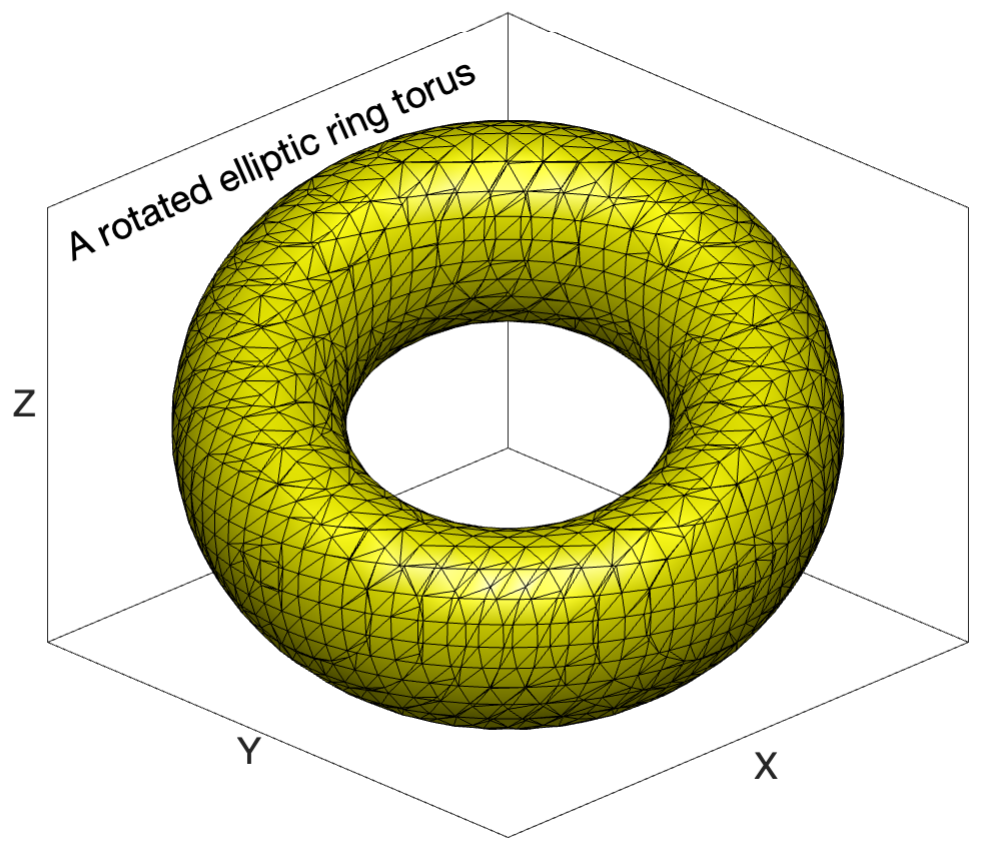} \;\;
\includegraphics[width=3.0in]{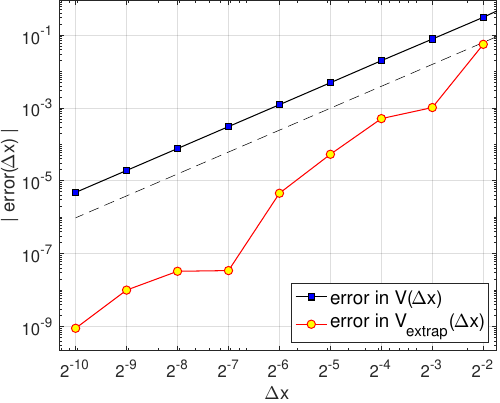}
\end{center}
\vskip -0.75cm
\caption{Left: Triangulation of the rotated elliptic ring torus as an isosurface of 
$f(\mathbf{x})$ in \eqref{prob_6}. Right: errors in numerical volumes $V(\Delta x)$ and 
$V_\text{extrap}(\Delta x)$ vs mesh spacing $\Delta x$.} 
\label{fig_06}
\end{figure}

\begin{equation}
\begin{dcases}
\text{function: } f(\mathbf{x}) = r^2-\Big(
\Big[ \sqrt{\tilde{x}^2+\tilde{y}^2} - R\Big]^2+\Big(\frac{\tilde{z}}{c}\Big)^2
 \Big), \quad  \mathbf{x} = (x, y, z) \\[0ex] 
\qquad \text{where }
\begin{pmatrix} \tilde{x} \\ \tilde{y} \\ \tilde{z} \end{pmatrix}
= B \begin{pmatrix} x \\ y \\ z \end{pmatrix}, \quad 
B = \begin{pmatrix} \frac{\sqrt{3}}{2\sqrt{2} } & \frac{\sqrt{3}}{2\sqrt{2} } & \frac{1}{2} \\[1ex]
\frac{-1}{\sqrt{2}} & \frac{1}{\sqrt{2}} & 0 \\ 
\frac{-1}{2\sqrt{2}} & \frac{-1}{2\sqrt{2}} & \frac{\sqrt{3}}{2} \end{pmatrix}
 \\[0ex]
\qquad \qquad R = 0.8, \;\; r = 0.28, \;\; 
c = \frac{1}{0.7} \\[1ex]
\text{enclosed region: } D = \{(x, y, z) \big| f(\mathbf{x}) \ge 0 \} \\[1ex]
\text{volume}(D) = 2 \pi^2 c\, r^2 R  
\end{dcases}
\label{prob_6}
\end{equation}
In \eqref{prob_6}, the orthogonal transformation represented by matrix $B$ is 
different from that by matrix A in \eqref{prob_2} and \eqref{prob_4}. 
If we first rotate the elliptic ring torus about the $z$-axis, due to the axial symmetry of the object, 
this rotation would not change the object at all. 
Instead, in \eqref{prob_6}, transformation represented by matrix $B$ 
corresponds to rotating the 3D object 
first with respect to the $y$-axis by $(-30^\circ)$ and then rotating it with respect to the 
$z$-axis by $45^\circ$. 
A triangulation of the rotated elliptic ring torus is shown in the left panel of Figure \eqref{fig_06}. 
The error trends of numerical volumes are 
plotted in the right panel of Figure \eqref{fig_06}. 
Similar to what we observed in all other test problems, 
the error in $V(\Delta x)$ is proportional to $(\Delta x)^2$; the error in 
$V_\text{extrap}(\Delta x)$ is significantly below that in $V(\Delta x)$. 
\FloatBarrier
%

\section{A Simplified Case of Electromagnetic Heating}
We study the electromagnetic heating of a flat skin by an incident electromagnetic wave of 
millimeter wave length. We consider the idealized situation where the incident wave 
is axially symmetric and is perpendicular to the flat skin surface. 
We first establish the coordinate system as described in our previous study \cite{WBZ_2020, Wang_2020}. 
The normal direction of the flat skin surface pointing into the skin tissue is selected as 
the positive $z$-direction. The skin surface is at $z=0$, and $z > 0$ corresponds 
to the skin tissue beneath the surface.
The center of the incident beam is selected as the origin of the $(x, y)$-plane. 
In the skin tissue, the 3D coordinates of a point are represented by 
$(x, y, z), \; z \ge 0$. 
\begin{figure}[!h]
\vskip 0.0cm
\begin{center}
\includegraphics[height=3.0in]{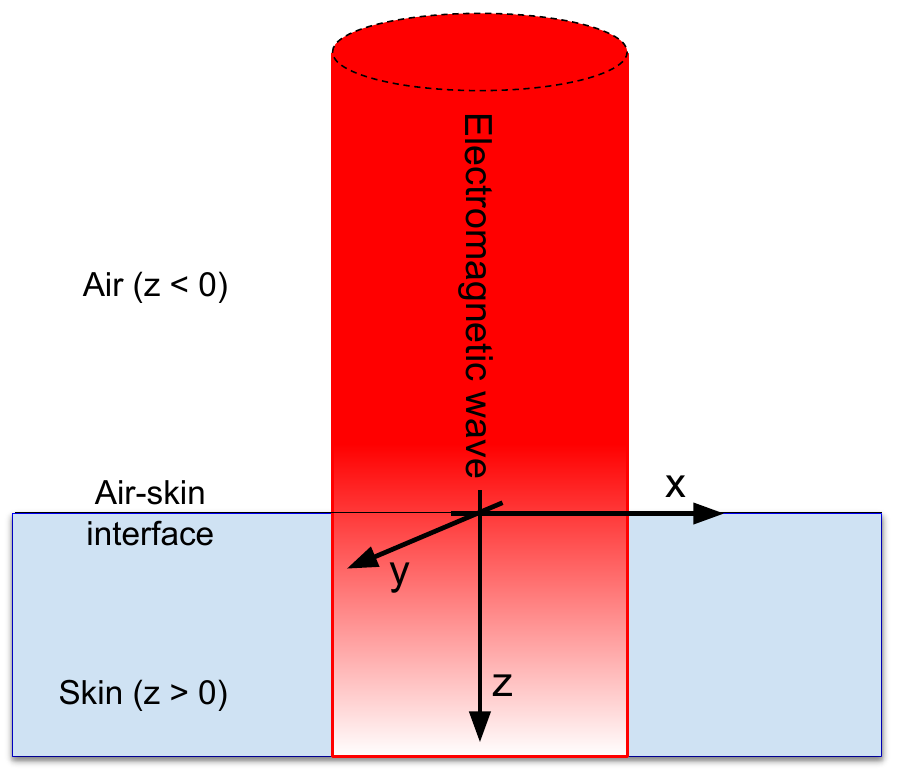} 
\end{center}
\vskip -0.75cm
\caption{The coordinate system for skin and incident 
electromagnetic wave.} 
\label{fig_beam_skin}
\end{figure}
\FloatBarrier

\subsection{Physical quantities, equations, and non-dimensionalization}
We introduce the physical quantities and equations used in our model. 
\begin{itemize}
\item $T(x, y, z, t)$ is the skin temperature at position $(x, y, z)$ at time $t$.
\item $\rho_m$ is the mass density, $C_p$ the specific heat capacity, and 
$k$ the heat conductivity of skin. 
In this study, we consider a single layer of uniform skin where 
all material properties are independent of position $(x, y, z)$.
\item
The energy balance gives the governing equation for $T(x, y, z, t)$ \cite{Wang_2025}:
\begin{equation}
\underbrace{\rho_m C_p\frac{\partial T}{\partial t}}_{\substack{
\text{rate of change}\\ \text{of heat in skin}}}
= \underbrace{k \Big(\frac{\partial^2T}{\partial x^2}
+\frac{\partial^2T}{\partial y^2}+\frac{\partial^2T}{\partial z^2}
\Big)}_{\substack{
\text{net heat influx}\\ \text{from conduction}}}
+ \underbrace{P_d(x, y)\mu e^{-\mu z}}_{\substack{\text{absorbed power}\\ 
\text{as heat source}}}
\label{heat_Eq1}
\end{equation}
\item $\mu$ is the absorption coefficient of skin for the
electromagnetic frequency as the wave propagates in the skin. 
$\mu $ describes the fraction of power absorbed per unit depth
as the beam propagates in the skin, which is a lossy medium. 
The remaining power density at depth $z$ is governed by Beer–Lambert law \cite{Beer-Lambert}.
\begin{equation}
\text{(remaining power density at depth $z$)} \propto e^{-\mu z}
\label{remaining_P}
\end{equation}
$\mu$ has the physical dimension of $1/[\text{length}]$. 
$(1/\mu)$ gives the characteristic penetration depth of the electromagnetic frequency, 
the depth at which the power density is attenuated by a factor of $1/e$. 
$(1/\mu)$ serves as the depth scale in non-dimensionalization. 
\item $P_d(x, y)$ is the power density passing through the surface 
into skin tissue. Since skin is a lossy medium ($\mu > 0$), 
as described in Beer–Lambert law in \eqref{remaining_P}, the remaining power density 
at depth $z$ is virtually zero at large $(\mu z)$. 
All of $P_d(x, y)$ is converted to heat during the electromagnetic propagation in skin. 
The heat production rate per volume at depth $z$ is $P_d(x, y)\mu e^{-\mu z}$, which 
is the heat source in the skin temperature evolution governed by 
the 3D heat equation \eqref{heat_Eq1}. 
\item 
In our simple model, the power density $P_d(x, y)$ is an axially symmetric Gaussian. 
\[ P_d(x, y) = P_d^{(0)}\exp(\frac{-2(x^2+y^2)}{w^2}) 
= P_d^{(0)}\exp(\frac{-(x^2+y^2)}{2 \sigma^2}) \]
where $w$ is the standard beam radius defined as the distance from the beam center to 
the location where the power density drops to $1/e^2 \approx 13.5\%$ of the peak value; 
$\sigma $ is the RMS (root mean square) beam radius, which is 
the standard deviation of the Gaussian distribution. These two are related by 
$\sigma = w/2 $. 
\item 
In the $z$-direction, the characteristic scale is given by the penetration depth $z_s = 1/\mu$. 
In the $(x, y)$-lateral directions, the characteristic scale is given by the RMS beam radius 
$r_s = \sigma$. In the time direction, the characteristic scale is given by the quantity 
$t_s \equiv \frac{\rho_m C_p}{k \mu^2}$. 
With these scales, we carry out non-dimensionalization and recycle the same notations 
for all non-dimensional functions and variables afterwards. 
The resulting non-dimensional governing equation is : 
\begin{equation}
\underbrace{\frac{\partial T}{\partial t}
= \frac{\partial^2T}{\partial z^2} + \varepsilon^2 \Big(\frac{\partial^2T}{\partial x^2}
+\frac{\partial^2T}{\partial y^2}\Big)
+ P_d^{(0)} \exp(\frac{-(x^2+y^2)}{2}) e^{-z}
}_\text{governing equation after non-dimensionalization}
\label{heat_Eq1B}
\end{equation}
where $\varepsilon \equiv \frac{z_s}{r_s}$ is the ratio of the penetration depth to 
the lateral scale \cite{Jaime-Yepez_2024}. 
\item 
In electromagnetic heating applications using millimeter wavelength, the electromagnetic 
penetration depth is sub-millimeter while the RMS radius of the Gaussian beam is 
several centimeters \cite{Walters_2000}. In these applications, the depth to lateral length ratio 
satisfies $\varepsilon \ll 1$. Thus, the effect of the lateral 
heat conduction $\varepsilon^2 \Big(\frac{\partial^2T}{\partial x^2}
+\frac{\partial^2T}{\partial y^2}\Big)$ is negligible in comparison with 
other terms in \eqref{heat_Eq1B}. 
To the leading order, the governing equation becomes: 
\begin{equation}
\underbrace{\;\; \frac{\partial T}{\partial t}
= \frac{\partial^2T}{\partial z^2}+ P_d^{(0)} \exp(\frac{-(x^2+y^2)}{2}) e^{-z}
\;\;}_\text{simplified non-dimensional equation}
\label{heat_Eq1C}
\end{equation}
\end{itemize}

\subsection{Initial boundary value problem and its analytical solution}
We continue discussing the simplified system governing the skin temperature evolution. 
\begin{itemize}
\item 
Before the start of electromagnetic exposure, skin is in its thermal homeostasis. 
The homeostatic temperature varies gradually from the core temperature inside skin tissue
to a lower temperature at skin surface. 
In out simplified model, however, we neglect this spatial variation in homeostatic temperature
and assume that skin baseline temperature is a uniform constant $T_\text{base}^\text{(phy)}$ 
at the beginning of electromagnetic exposure. 
This assumption is justified for millimeter wave exposure 
because the heating is concentrated in a very thin skin layer in which the variation of 
homeostatic temperature is small. This assumption leads to a simple initial condition. 
\[ T^\text{(phy)}(x, y, z, 0) = T_\text{base}^\text{(phy)} \] 
\item
In non-dimensionalzation \eqref{heat_Eq1B}, physical temperature $T^\text{(phy)}$ is 
shifted by $T_\text{base}^\text{(phy)}$ and normalized by the temperature scale 
$\Delta T \equiv T_\text{act}^\text{(phy)}-T_\text{base}^\text{(phy)}$
where  $T_\text{act}^\text{(phy)}$ is the physical temperature threshold 
for activating skin thermal nociceptors. 
Given the physical baseline temperature $T_\text{base}^\text{(phy)}$, temperature scale 
$\Delta T$ is the temperature increase needed for skin thermal nociceptor activation. 
After normalization, the non-dimensional versions of baseline temperature and temperature threshold are 
\[ T_\text{base} = 0, \qquad T_\text{act} = 1\] 
\item 
In homeostasis, skin loses heat to the environment at the skin 
surface via black body radiation and air convection cooling. 
However, the rate of skin surface heat loss is low. 
In a high-intensity short-duration electromagnetic exposure, 
the rate of electromagnetic energy passing into the skin tissue is much larger than the rate 
of slow heat loss at skin surface. 
Thus, we neglect the slow heat loss at the skin surface
and assume that there is no heat exchange with the environment at the skin surface.
This assumption leads to a simple boundary condition.  
\[ \frac{\partial T(x, y, z, t)}{\partial z}\Big|_{z=0} = 0 \] 
\item 
The governing system for the skin temperature evolution consists of 
the simplified non-dimensional equation and initial and boundary conditions. 
\begin{equation}
\begin{dcases}
\frac{\partial T}{\partial t}
= \frac{\partial^2T}{\partial z^2}+ P_d^{(0)} \exp(\frac{-(x^2+y^2)}{2}) e^{-z} \\[2ex]
\frac{\partial T(x, y, z, t)}{\partial z}\Big|_{z=0} = 0 \\[1ex]
T(x, y, z, 0) = 0
\end{dcases}
\label{T_IBVP}
\end{equation}
\item 
Note that \eqref{T_IBVP} has three key features: i) it is linear; 
ii) $(x, y)$ are not differential variables; and 
iii) the source term has separate dependences on $(x, y)$ and on $z$. 
These features imply that the skin temperature $T(x, y, z, t)$ has the form 
\begin{equation}
T(x, y, z, t) = \underbrace{P_d^{(0)} \exp(\frac{-(x^2+y^2)}{2})}_{
\text{$(x, y)$-dependence}} \underbrace{\; U(z, t) \;}_{\text{$z$-dependence}}
\label{T_Pd_U}
\end{equation}
where $U(z, t) $ is a function of $(z, t)$ only and satisfies 
\begin{equation}
\begin{dcases}
\frac{\partial U}{\partial t}
= \frac{\partial^2 U}{\partial z^2}+ e^{-z} \\[2ex]
\frac{\partial U(z, t)}{\partial z}\Big|_{z=0} = 0 \\[1ex]
U(z, 0) = 0
\end{dcases}
\label{U_IBVP}
\end{equation}
\item The solution of IBVP \eqref{U_IBVP} has a closed form analytical expression
\cite{WBZ_2020C}. 
\begin{equation}
U(z, t) = \begin{dcases} 
\frac{e^{-z+t}}{2} \text{erfc}(\frac{-z+2t}{\sqrt{4t} }) 
+\frac{e^{z+t}}{2} \text{erfc}(\frac{z+2t}{\sqrt{4t} }) & \\
\hspace{1cm} -e^{-z}-z \, \text{erfc}\Big(\frac{z}{\sqrt{4 t} }\Big) 
+ \frac{2\sqrt{t}}{\sqrt{\pi }} e^{\frac{-z^2}{4t}}, & t > 0 \\
0, & t \le 0 
\end{dcases}. 
\label{U_sol} 
\end{equation}
where $\text{erfc}(s)$ is the complementary error function.
\[ \text{erfc}(s) = \frac{2}{\sqrt{\pi}} \int_s^{+\infty} \exp(-t^2) dt \] 
\end{itemize}

\subsection{A test problem for skin activated volume}
\begin{figure}[!h]
\vskip 0.0cm
\begin{center}
\includegraphics[height=2.5in]{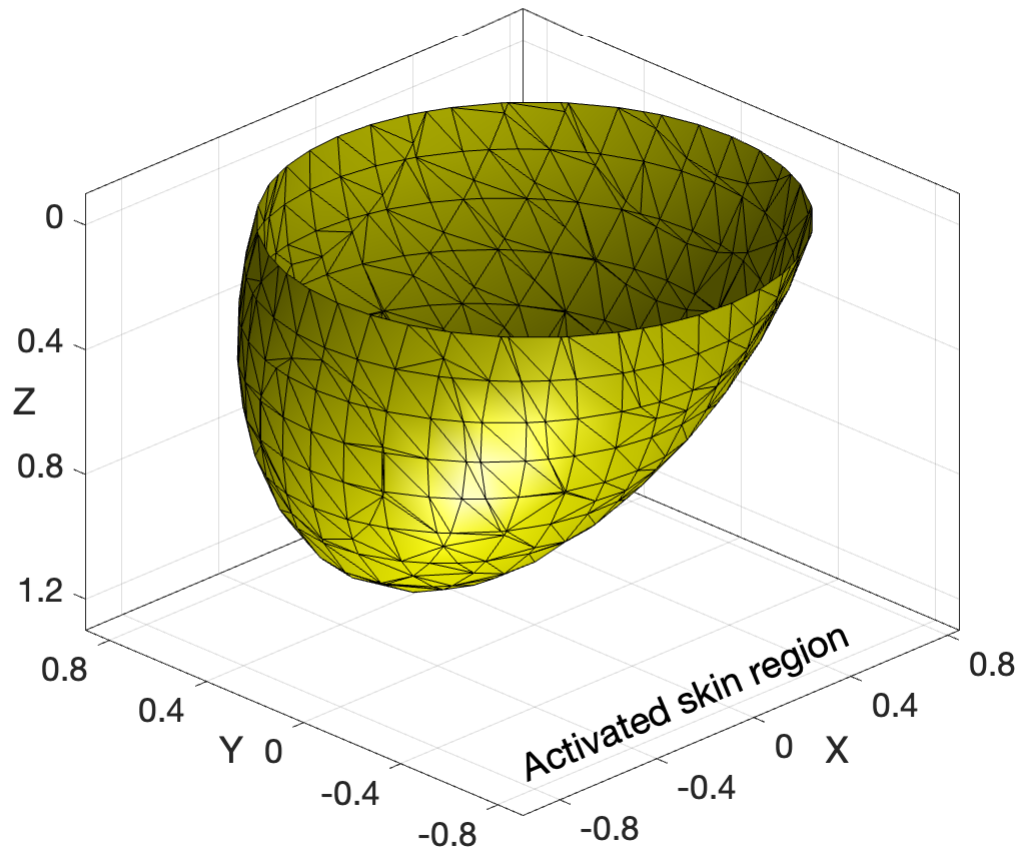} \;\;
\includegraphics[width=3.0in]{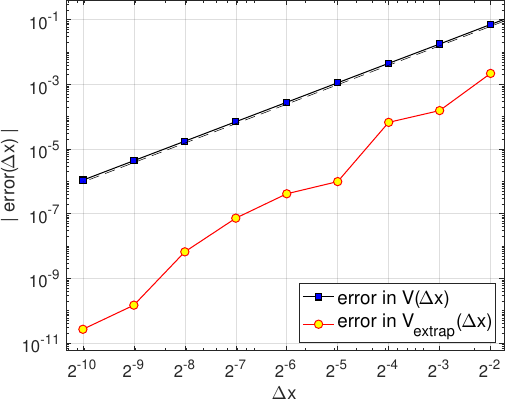}
\end{center}
\vskip -0.75cm
\caption{Left: Triangulation of the activated skin region as an isosurface of 
$f(\mathbf{x})$ in \eqref{prob_7}. Right: errors in numerical volumes $V(\Delta x)$ and 
$V_\text{extra}(\Delta x)$ vs mesh spacing $\Delta x$.} 
\label{fig_07}
\end{figure}

In this subsection, all quantities are non-dimensional. 
The activated skin region $D$ at time $t$ is enclosed by the isosurface of skin temperature 
$T(x, y, z, t)$ at value $T_\text{act} = 1$. 
Using the expression of $T(x, y, z, t)$ in \eqref{T_Pd_U}, 
we write the activated skin region $D$ as 
\begin{equation}
\begin{dcases}
\text{function: } f(\mathbf{x}) = P_d^{(0)} \exp(\frac{-(x^2+y^2)}{2}) U(z, t)-1, 
\qquad  \mathbf{x} = (x, y, z) \\[1ex] 
\text{enclosed region: } D = \{(x, y, z) \big| f(\mathbf{x}) \ge 0 \} \end{dcases}
\label{D_heating}
\end{equation}
In this simplified electromagnetic heating problem, the activated region $D$ is axially symmetric. 
In more realistic electromagnetic heating applications, the electromagnetic beam may be moving 
relative to skin, the incident angle may be oblique, or the skin material properties 
may be inhomogeneous. 
Under these complex conditions, we no longer have a close form analytical solution for 
skin temperature, which needs to be solved numerically on a 3D grid 
and in general is not axially symmetric \cite{JOSHI2022108506, dynamics5040041}. 
Our volume computation methods are well capable of accommodating this complex situation. 
It requires only a discrete representation of skin temperature on a 3D grid; no symmetry 
is assumed. 
Here we use \eqref{D_heating} to construct a non-symmetric problem 
for testing accuracy of volume computation. 
We first find the activated volume of \eqref{D_heating}. 
The activated skin region of \eqref{D_heating} is  
\begin{align}
& D = \big\{(x, y, z) \, \Big| \; P_d^{(0)} \exp(\frac{-(x^2+y^2)}{2}) U(z, t) \ge 1 \big\} 
\nonumber \\[1ex]
& \quad = \Big\{(x, y, z) \, \Big| \; (x^2+y^2) \le 2 \log\big(P_d^{(0)} U(z, t) \big), 
\; 0 \le z \le z_L(t) \Big\} 
\label{D_exp2}
\end{align}
where $z_L(t) \equiv U^{-1}\big(\frac{1}{P_d^{(0)}} , t \big)$ and $U^{-1}(\cdot , t)$ 
is the inverse of $z \mapsto U(z, t)$ at fixed $t$. 
We express the activated skin volume in terms of a single integral. 
\[ \text{volume}(D) = 2\pi \int_0^{z_L(t)} \log\big(P_d^{(0)} U(z, t) \big) d z \]
We solve this integral accurately to machine precision
using the composite Simpson's rule, and use the result to assess the errors in numerical volumes 
\eqref{V_method} and \eqref{V_extrap}. 
In the numerical test, we use $P_d^{(0)} = 2.5$ and $t = 1$. 
To test the robustness of the volume computation, we shift the region in the $x$-direction 
by a $z$-dependent amount. This transformation destroys the axial symmetry while 
preserving the true volume, which is needed in error assessment. 
We set the test problem as follows. 
\begin{equation}
\begin{dcases}
\text{function: } f(\mathbf{x}) = P_d^{(0)} \exp(\frac{-((x+0.5z)^2+y^2)}{2}) U(z, t)-1, 
\quad  \mathbf{x} = (x, y, z) \\[1ex] 
\text{enclosed region: } D = \{(x, y, z) \big| f(\mathbf{x}) \ge 0 \} \\[1ex]
\text{volume}(D) = 1.29409605751414 \text{ (at $P_d^{(0)} = 2.5$, $t = 1$)}
\end{dcases} 
\label{prob_7}
\end{equation}
Figure \ref{fig_07} shows a triangulation of the isosurface at the activation temperature 
threshold (left panel) and the errors in numerical volumes (right panel). 
The triangulation is based on a coarse grid with 
$\Delta x = \frac{1}{8}$ for illustration purpose. 
The activated skin region $D$ is enclosed by the isosurface and the skin surface ($z=0$). 
With the $z$-dependent shift in $x$, neither function $f(\mathbf{x})$ in \eqref{prob_7} nor 
the region $D$ enclosed by the isosurface is axially symmetric. 
The volume computation procedure, however, is not impacted at all by the transformation. 
It uses only a discretization of function $f(\mathbf{x})$ on a 3D numerical grid. 
The errors in numerical volumes behave as expected. 
The error in $V(\Delta x)$ is proportional to $(\Delta x)^2$; the error in 
$V_\text{extrap}(\Delta x)$ is significantly below that in $V(\Delta x)$. 
\FloatBarrier
%

\section{Summary}
In this study, we investigated an accurate and reliable method for computing the activated 
skin volume in electromagnetic heating when skin is exposed to an incident electromagnetic 
beam of millimeter wavelength. 
The activated volume is enclosed by the isosurface of skin temperature at the level of thermal 
nociceptor activation threshold. Thus, the task of computing the activated volume is cast 
into a general problem of finding the volume enclosed by an isosurface 
of an underlying function $f(\mathbf{x})$. 
Mathematically, the enclosed volume is expressed as a surface integral 
over the isosurface. 

In many electromagnetic heating applications, the skin temperature is 
obtained numerically by solving a heat transfer model on a 3D grid. That is, 
the skin temperature is known only on a 3D grid. 
With only a discrete representation of function $f(\mathbf{x})$, 
the corresponding isosurface is obtained approximately using linear interpolation 
along grid lines of the 3D grid. The resulting approximation of the isosurface 
is a collection of planar triangle elements. 
With the isosurface being approximately represented by a triangulation, 
surface integration is carried out over each planar triangle element, which 
is analytically given by a triple scalar product. 
The enclosed volume is obtained by summing over all triple scalar products, each 
for one planar triangle element of the isosurface. 

The process of approximating a surface using flat triangle elements is analogous to 
that of approximating a curve using line segments. 
Both are expected to produce second order approximations. 
Given a discrete representation of $f(\mathbf{x})$ with mesh spacing $\Delta x$, 
a triangulation of the isosurface is constructed and the enclosed volume is calculated. 
The dominant error in the numerical volume obtained with mesh spacing $\Delta x$ 
is proportional to $(\Delta x)^2$. 
From the given fine representation of $f(\mathbf{x})$, a coarse representation
with mesh spacing $(2 \Delta x)$ is obtained by down sampling, 
which does not involve any additional numerical simulations of 3D heat transfer model. 
The dominant error in the numerical volume obtained with mesh spacing $(2\Delta x)$ 
is proportional to $(2 \Delta x)^2$. 
Extrapolation combines these two numerical volumes to eliminate the 
dominant error. The result of extrapolation is a more accurate approximation to 
the enclosed volume. 

We evaluated the performance of the two volume computation methods: i) triangulation only, 
ii) triangulation plus extrapolation. 
To accurately assess the errors in numerical volumes, we selected test problems 
that have closed form analytical solutions or solutions, including a simplified case of  
electromagnetic heating. 
In all problems tested, we observed that i) the error in 
the numerical volume directly from triangulation of isosurface is proportional to $(\Delta x)^2$
where $\Delta x$ is the mesh spacing of the 3D grid in the given discrete representation of 
function $f(\mathbf{x})$; 
and ii) the error in the numerical volume from triangulation plus extrapolation 
is much smaller than that in the triangulation-only volume, 
clearly demonstrating that extrapolation consistently improves the accuracy
across all test cases.

\vskip 1in
\noindent{\bf \large Acknowledgement and disclaimer}

\noindent \indent
The authors acknowledge the Joint Intermediate Force Capabilities Office of U.S. Department of Defense and the Naval Postgraduate School for supporting this work. The views expressed in this document are those of the authors and do not reflect the official policy or position of the Department of Defense or the U.S. Government.


%

\vskip 1in
\begin{thebibliography}{AA}
%
\bibitem{Romanenko_2017}
Romanenko, S., R. Begley, A. R. Harvey, L. Hool, and V. P. Wallace (2017) The interaction between electromagnetic fields at megahertz, gigahertz and terahertz frequencies with cells, tissues and organisms: risks and potentials. Journal of the Royal Society Interface 14: 20170585 

\bibitem{Zhadobov_2011}
Zhadobov, M., N. Chahat, R. Sauleau, C. Le Quement, and Y. Le Drean (2011) Millimeter-wave interactions with the human body: state of knowledge and recent advances. International Journal of Microwave and Wireless Technologies 3:237-247 

\bibitem{Quement_2014}
Le Quement, C., C. N. Nicolaz, D. Habauzit, M. Zhadobov, R. Sauleau, and Y. Le Drean (2014) Impact of 60-GHz millimeter waves and corresponding heat effect on   endoplasmic reticulum stress sensor gene expression. Bioelectromagnetics 35:444-451 

\bibitem{Nelson_2000}
Nelson, D.A., M. T. Nelson, T. J. Walters and P. A. Mason (2000) Skin heating effects of millimeter-wave irradiation – thermal modeling results. IEEE Transactions on Microwave Theory and Techniques 48:2111-2120 

\bibitem{Foster_2010}
Foster, K. R., H. Zhang and J. M. Osepchuk (2010) Thermal response of tissues to millimeter waves: implications for setting exposure guidelines. Health physics 99(6):806-810 

%
\bibitem{Cazares_2019}
Cazares, S.M., Snyder, J.A., Belanich, J., Biddle, J.C., Buytendyk, A.M., 
Teng, S.H.M. and O’Connor, K. (2019) 
Active Denial Technology Computational Human Effects End-to-End Hypermodel 
for Effectiveness (ADT CHEETEH-E). 
{\em Human Factors and Mechanical Engineering for Defense and Safety}, 
{\bf 3}, Article No. 13. 
doi:10.1007/s41314-019-0023-7

\bibitem{Topfer_2015}
Topfer, F. and Oberhammer, J. (2015)
Millimeter-wave Tissue Diagnosis: The most Promising Fields for Medical Applications.
{\em IEEE Microwave Magazine}, {\bf 16}(4), 97-113.

\bibitem{Sri_2020}
Srivastave, A. (2020) A. Srivastave, Energy efficient transmission trends towards future green cognitive radio networks (5G): Progress, taxonomy and open challenges, Journal of Network and Computer Applications 168: 102760

\bibitem{Sheen_2007}
Sheen, D.M., , D. L. McMakin and T. E. Hall (2007) Detection of explosives by millimeter-wave imaging, in Counterterrorist Detection Techniques of Explosives, edited by J. Yinon. 237-278

\bibitem{Liu_2020}
Liu, H. (2020)  Robot Systems for Rail Transit Applications. Elsevier.

\bibitem{WBZ_2020}
Wang, H., W. Burgei, W.A. and H. Zhou (2020) 
A concise model and analysis for heat-induced withdrawal reflex caused by 
millimeter wave radiation.
American Journal of Operations Research, 10, 31-81. 

\bibitem{Wang_2020}
Wang, H. ,W. Burgei, W. and H. Zhou (2020) 
Non-Dimensional Analysis of Thermal Effect on Skin Exposure to 
an Electromagnetic Beam. American Journal of Operations Research, 
10, 147-162. 

\bibitem{Wang_2025}
H. Wang, S. E. Foley, and H. Zhou (2025) Asymptotic Solution for Skin Heating by an Electromagnetic Beam at an Incident Angle, \textit{Electronics}, vol. 14, no. 15, p. 3061.

\bibitem{Beer-Lambert}
Skoog, D.A., Holler, F.J. and Crouch, S.R. (2017) 
Principal of Instrumental Analysis. 7th Edition, Sunder College Publisher, New York.

\bibitem{Jaime-Yepez_2024}
Jaime-Yepez, U., Wang, H., Foley, S. E., and Zhou, H. (2024) Asymptotic solution of electromagnetic heating of skin tissue with lateral heat conduction. {\em J Eng Math}, {\bf 147}, 14. 
doi: doi.org/10.1007/s10665-024-10390-y

\bibitem{Walters_2000}
Walters T.J., Blick D.W., Johnson L.R., Adair E.R., Foster K.R. (2000) 
Heating and Pain Sensation Produced in Human Skin by Millimeter Waves: Comparison to a Simple Thermal Model. {\em Health Physics}, {\bf 78}(3): 259–267.
doi:10.1097/00004032-200003000-00003

\bibitem{WBZ_2020C}
Wang, H., Burgei, W.A. and Zhou, H. (2020) 
Analytical solution of one-dimensional Pennes' bioheat equation. 
{\em Open Physics}, {\bf 18}(1), 1084-1092.
doi: 10.1515/phys-2020-0197

\bibitem{dynamics5040041}
A.~S. Rodríguez-Pérez, H.~E. Gilardi-Velázquez, and S.~E. Velázquez-Pérez (2025)
Analysis and Simulation of Dynamic Heat Transfer and Thermal Distribution in Burns with Multilayer Models Using Finite Volumes,
\emph{Dynamics}, vol.~5, no.~4, Art.~41.
doi:10.3390/dynamics5040041.

\bibitem{JOSHI2022108506}
A.~Joshi, F.~Wang, Z.~Kang, B.~Yang, and D.~Zhao (2022)
A three-dimensional thermoregulatory model for predicting human thermophysiological responses in various thermal environments,
\emph{Building and Environment}, vol.~207, Art.~108506.
doi:10.1016/j.buildenv.2021.108506.

%
%
%
%
%
%
%
\end{thebibliography}
\end{document}